# Classical Localization of an Unbound Particle in a Two-Dimensional Periodic Potential and Surface Diffusion

Roger Haydock

Department of Physics and Materials Science Institute

University of Oregon, Eugene OR 97403-1274, USA

**Abstract**

In periodic, two-dimensional potentials a classical particle might be expected to escape from any finite region if it has enough energy to escape from a single cell. However, for a class of sinusoidal potentials in which the barriers between neighboring cells can be varied, numerical tridiagonalization of Liouville's equation for the evolution of functions on phase space reveals a transition from localized to delocalized motion at a total energy significantly above that needed to escape from a single cell. It is argued that this purely elastic phenomenon increases the effective barrier for diffusion of atoms on crystalline surfaces and changes its temperature dependence at low temperatures when inelastic events are rare.

1.  A Liouvillian Approach to Classical Mechanics

The classical dynamics of solids is difficult to calculate because the forces are non-linear, the many degrees of freedom are strongly coupled, and the time-scales for macroscopic motion diverge from the vibrational periods of individual atoms. In this work, recent developments in the solution of Liouville's equation [1,2] are applied numerically to motion in a two-dimensional elastic potential. The use of the Liouvillian operator to evolve functions rather than Newton's equations to evolve trajectories has the advantages that functions sample all of phase space rather than the infinitesimal fraction of it sampled by any finite number of trajectories, and that Liouville's equation is linear in contrast with the non-linear equations for trajectories. Because of this linearity the problem can be attacked with a variety of reliable analytic and numerical methods.

The methods applied in this work are based on the recursion method [3] for calculating projected densities of states in quantum mechanics. More generally, the recursion method is a way of calculating projected power spectra for linear wave equations of which Liouville's equation in classical mechanics is an example, see further Sec. 2. These methods have previously been applied to one-dimensional systems [2], and to calculations of relaxation rates for some two-dimensional systems [1]. While this approach is not limited to low dimensionality or few particles, the work presented here is part of a systematic exploration of classical dynamics in solids, in this case addressing the motion of adatoms on a crystalline surface.

The first step in these calculations is tridiagonalization of Liouville's equation. Although tridiagonalization has a long history in physics and mathematics, a thorough understanding of the mathematical properties of tridiagonalization was achieved in the context of the classical moment problem [4]. Some noteworthy applications of tridiagonalization are the Lanczos method for solving the eigenproblem [5], Zwanzig's work in classical statistical mechanics [6], Mori's method for quantum statistical mechanics [7], and Risken's continued fraction method [8] for the Fokker-Planck equation. Recently Sinkovits and others [9] have applied tridiagonalization to anharmonic oscillators. The work presented in this paper differs from related work in statistical mechanics in that energy is consved because there is no thermal averaging and consequently no particular temperature. This work also differs from related work in classical mechanics in that all forces are non-singular, resulting in convergent continued fractions.

The main result presented here is evidence of a localization transition for a particle in a two-dimensional, periodic potential at a total energy well above the barrier between cells. This is surprising from the quantum perspective where all states are delocalized in periodic potentials according to Bloch's theorem [10]. While classical systems are known to undergo percolation transitions[11] when a random array of obstacles becomes so dense that there is no path through, the transition found here takes place for a system in which the scatterers are periodic rather than random, and at a density of scatterers for which there is plenty of room for the particle to get between them. Periodic, two-dimensional potentials have been studied previously using the Fokker-Planck equation, see for example Ref. 12, however the friction term and thermal averaging in that approach obscure the localization which is the focus of this paper.

The remainder of the paper consists of three Secs. of which the first briefly reviews the application of the recursion method to classical dynamics, and then shows how macroscopic limits may be taken in time and distance. Section 3 presents the model for a particle bound to an elastic surface and the results which characterize its evolution. Finally, Sec. 4 relates the results of Sec. 3 to the diffusion of adatoms having weak inelastic interactions with a surface, and suggests ways these effects might be observed experimentally.

## 2. Recursion Method for Classical Dynamics

The work presented here builds on that described in Haydock and Kim [2] and Haydock, Nex, and Simons [1] which are briefly reviewed. Calculation of macroscopic properties such as diffusion constants require evaluation of limits which are discussed in Sec. 2.4.

### 2.1 Classical Wave Functions

Newton's equations of motion for a system of particles combine into an ordinary differential equation,

$$dX/dt = F(X), \qquad (1)$$

where X is a vector whose components are the positions and momenta of the particles, and the corresponding components of F(X) are the velocities and accelerations of the particles. In general, F depends non-linearly on X, so Eq. 1 is difficult to solve. However, a function $u(X;t)$, whose values propagate along the trajectories which satisfy Eq. 1, obeys a linear partial differential equation obtained by setting the total time-derivative of the function to zero:

$$i \, \partial u(X; t)/\partial t = L \, u(X; t), \qquad (2)$$

where $L$ is the Liouvillian operator,

$$L = -i \, F(X) \cdot \nabla_X, \qquad (3)$$

and $\nabla_X$ is the gradient operator for functions on phase space. When u(X;t) is a density (non-negative) in phase space, Eq. 2 is Liouville's equation. The roots of -1 are introduced in Eqs. 2 and 3 so that the Liouvillian operator $L$ is Hermitian with respect to integration over all of phase space provided that $\nabla_X \cdot F(X)$ is zero.

Because Eq. 2 is a wave equation and because its solutions can be multiplied to give other solutions, it is shown below that this equation serves as a classical analogue of the Schrödinger equation from quantum mechanics. The wave property of Eq. 2 arises because its solutions can be written as superpositions of the products of functions $\psi_\omega(X)$ on phase space and purely harmonic complex phases $e^{i\omega t}$, where the eigenvalue $\omega$ of $L$ is real because $L$ is Hermitian. Since F(X) is real, $\psi_\omega(X)^*$ is an eigenfunction of $L$ with eigenvalue $-\omega$, and since $L$ is a linear combination of first derivatives, products of eigenfunctions are also eigenfunctions whose eigenvalues are the sum of the eigenvalues of the factors. The consequence of these properties is that $|\psi_\omega(X)|^2$ is a stationary density just as in quantum mechanics. As with Schrödinger's equation, a complete basis for solutions to Eq. 2 can be constructed from eigenfunctions of the Liouvillian. It is particularly convenient to formulate Liouvillian mechanics in terms of waves rather than distributions because no further constraints are necessary to maintain the non-negativity of distributions which are the squared magnitudes of waves. Since the waves are unconstrained, the problem may be attacked with the full power of linear algebra and linear operators. The only disadvantage of this approach is that like quantum wave functions, these classical wave functions have a gauge symmetry which is the choice of complex phase at each point in the mechanical phase space.

2.2 Tridiagonalization

While the system is defined by F(X), the states through which the system evolves are determined by u(X;0), the initial value of the function, from which Eq. 2 determines u(X;t). The method used here to solve Eq. 2 is tridiagonalization of the Liouvillian $L$, starting with a function $u_0(X)$ which is u(X;0) normalized so that the corresponding distribution (its squared magnitude) is normalized. The rest of the basis which tridiagonalizes $L$, $u_1(X)$, $u_2(X)$, ..., $u_n(X)$, ... is constructed by a Gram-Schmidt [13] recursion: acting on the last function with $L$, orthogonalizing the result to the previous function, and then normalizing its squared magnitude. The inner product used here for the orthogonalization and normalization is integration over phase space $\Gamma$,

$$<v(X), u(X)> = \int_\Gamma v(X)^* \, u(X) \, d\Gamma, \qquad (4)$$

and because $L$ is Hermitian with respect to this inner product, each new function need only be orthogonalized with respect to the two preceding functions in order to make it orthogonal to all preceding functions. Provided that F(X) in Eq. 3 is purely real, $L$ is a purely imaginary Hermitian operator, so its diagonal elements in the tridiagonal basis are all zero.

Given the first n+1 elements of the tridiagonal basis $u_0(X)$, $u_1(X)$, ..., $u_n(X)$ and the off-diagonal elements of the tridiagonal matrix $b_{-1}=0$, $b_0$, $b_1$, ..., $b_{n-1}$, the recursion is summarized in the following equations:

$$b_n = \sqrt{\langle (L\, u_n(X) - b_{n-1}\, u_{n-1}(X)), (L\, u_n(X) - b_{n-1}\, u_{n-1}(X))\rangle}, \quad (5)$$

$$u_{n+1}(X) = [(L\, u_n(X) - b_{n-1}\, u_{n-1}(X))]/ b_n. \quad (6)$$

Note that the numbering of the $\{b_n\}$ differs from previous work.

The tridiagonal matrix J is a one-dimensional projection of $L$ onto functions which span u(X;t): a projection because $\{u_n(X)\}$ does not contain functions which are inaccessible from u(X;0), and one-dimensional because u(X;0) contains at most one function from each subspace of functions with the same eigenvalue of $L$. While it is possible for $b_N$ to be zero for some finite N, in which case u(X,0) contains only N frequencies; in general J is an infinite matrix of which only a finite part can be computed numerically. The first few $u_n(X)$ describe short-time evolution, and the asymptotic (n goes to infinity) properties of $\{u_n(X)\}$ and $\{b_n\}$ describe the long-time behavior of u(X;t). In order to determine the long-time behavior of the system, $u_n(X)$ and $b_n$ must be calculated for sufficiently large n that the transient behavior associated with u(X;0) has been orthogonalized away (decayed away in the time domain) leaving the asymptotic behavior. It can be shown mathematically [4] that the transients decay exponentially with increasing n, so only a finite calculation is needed, but the size of that calculation depends on the details of the system, specifically the time it takes the system to explore various features in its structure.

2.3 The Power Spectrum of Fluctuations

The tridiagonal matrix is not itself an experimentally measurable quantity, so measurable quantities must be extracted from it to compare theory with experiment. One useful quantity which is easy to observe is the projected power spectrum, the squared magnitude of the Fourier transform of the component of u(X;0) in u(X;t), of fluctuations as the system evolves from u(X;0). This spectrum has sharp peaks at the frequencies with which the system tends to recur, and is smooth at frequencies for which it never recurs. In particular, the behavior of this power spectrum at zero frequency is related to the rate at which the initial state decays at long times, a diffusion constant.

It was shown in Ref. [1] that the projected power spectrum of fluctuations $g(\omega)$ is proportional to the imaginary part of the projected frequency resovent $R(\omega)$ for real $\omega$,

$$g(\omega) = |\text{Im}\{R(\omega)\}|/\pi, \tag{7}$$

and that $R(\omega)$ can be expanded as a continued fraction in the tridiagonal projection J of the Liouvillian,

$$R(\omega) = 1/(\omega - b_0^2/(\omega - b_1^2/(\omega - ... - b_n^2/(\omega - ...) ... ))). \tag{8}$$

This continued fraction expansion converges exponentially to the first sheet of $R(\omega)$ where there are no singularities except for $\omega$ real, and in previous work [1] this expansion was continued to the second sheet of $R(\omega)$ where there are singularities at complex frequencies corresponding to damped processes.

2.4 Macroscopic Limits

The simplest macroscopic motion is diffusion which is described by the diffusion constant D appearing in the diffusion equation,

$$\partial \rho(r; t)/\partial t = D \nabla^2 \rho(r; t), \tag{9}$$

where $\rho(r;t)$ is the density of atoms at r which only varies over macroscopic distances, t is time, and $\nabla^2$ is the Laplacian with respect to r. For a homogeneous, isotropic system Eq. 9 is satisfied by a decaying sinusoidal variation in density,

$$\rho(r; t) = 1 + \exp\{-i \Omega(k) t\} \sin k \cdot r, \tag{10}$$

where $\Omega(k)$ is a negative imaginary frequency, k is a macroscopic wave-number, and D is $i\Omega(k)/|k|^2$. Following Sec. 2.1, this density arises from the microscopic equations of motion as the squared magnitude of a superposition of two eigenfunctions of the Liouvillian, one which is constant over phase space with zero frequency and one whose average over momentum depends on position with wave-number k and has frequency $\Omega(k)$. Before averaging, the second eigenfunction can depend on the momentum of the adsorbed atom as well as its position, and might have components with wave-numbers other than k. However for the microscopic equations to be consistent with Eq. 10, wave-numbers other than k must disappear as k gets small and the second eigenfunction is averaged over momentum. More precisely, k is small when there are no correlations between momenta over distances comparable to $1/|k|$, and the result of this averaging must be a plane wave exp{ik·r}, so that the resulting density
|1+exp{i[k·r-$\Omega$(k)t]}|$^2$/2 is the same as in Eq. 10.

Bearing the above in mind, a simple approach to calculating the diffusion constant D is to tridiagonalize the Liouvillian starting with an initial function $u_0$(r,p) which is constant with respect to variation of momentum p on surfaces of constant energy, and which reduces to a plane wave exp{ik·r} in r when averaged over p. Since the Liouvillian operator is a gradient in phase space and there is no variation of $u_0$(r,p) in p, $Lu_0$(r,p) is [V(r,p)·k]$u_0$(r,p) where V(r,p) is the velocity vector as a function of r and p. From Eq. 5, $b_0$ is $\sqrt{\int[V(r,p)\cdot k]^2 drdp}$ which goes to zero for k small like $\beta$|k| because the velocity is isotropic. From Eq. 6, the next function $u_1$(r,p) in the tridiagonal basis is [V(r,p)·k]$u_0$(r,p)/|$\beta$k| whose variation with r is dominated by variations in the velocity due to the potential, and is almost independent of k. In the limit as k goes to zero, it is only $u_0$(r,p) and $b_0$ which depend on k, while the rest of the tridiagonalization, starting with $u_1$(r,p) is independent of k. Defining the k-independent tail of the continued fraction in Eq. 8 to be,

$$R_1(\omega) = 1/(\omega - b_1^2/(\omega - ... b_n^2/(\omega - ...) ... ))), \qquad (11)$$

the projected resolvent is,

$$R(\omega) = 1/[\omega - \beta^2 |k|^2 R_1(\omega)]. \qquad (12)$$

As argued above, the diffusion constant depends on the frequency of the eigenvalue of $L$ nearest zero. This eigenvalue appears in R($\omega$) as a pole on the second sheet, just below zero, or alternatively as a narrow peak in the imaginary part of R($\omega$) for real $\omega$ near zero. For k small, the power spectrum is almost completely concentrated in this peak which results from the single pole, so the pole is located on the second sheet of the negative imaginary axis at a distance from the origin given by the reciprocal of |R(0)|,

$$\Omega(k) = -i \beta^2 |k|^2 |R_1(0)|, \qquad (13)$$

and from this it follows by taking the limit in which k goes to zero that the diffusion constant is,

$$D = \beta^2 |R_1(0)|. \qquad (14)$$

From Eq. 11, the tail of the continued fraction $R_1(\omega)$ is an odd function of $\omega$, so $R_1(0)$ is purely imaginary.

According to Eq. 14 the projected resolvent determines the diffusion constant, and from previous work, see Ref. [14], it is the eigenvalues and eigenfunctions of the Liouvillian which determine the projected resolvent. In this case it is the eigenfunctions with eigenvalues near zero frequency which are important, and one of these is the constant function on phase space which is always an eigenfunction of the Liouvillian with zero frequency because its gradients are all zero. However, the constant function does not contribute to $R(\omega)$ for k not zero because $u_0(r,p)$ is orthogonal to the constant function. Using the above relations between eigenfunctions and resolvents, the diffusion constant is zero unless there is a band of eigenfunctions which are current-carrying in the sense that they belong to time-reversal doublets, and whose frequencies go to zero as $|k|^2$. Furthermore, if the frequency goes to zero other than as $|k|^2$, then the propagation is subdiffusive for powers greater than two and superdiffusive for powers less than two. If the eigenfunctions are non-current-carrying in the sense that they belong to time-reversal singlets, then the diffusion is anomalous.

3. Model for a Corrugated Surface

One of the simplest diffusing systems is an atom adsorbed on a crystalline surface. Except for hydrogen and helium, the motion of the adsorbed atom is well described by classical mechanics, and is complicated due to the non-linear forces between the adatom and the surface. In what follows this system is investigated by means of a simple model, a classical particle moving in an elastic potential for which the particle's speed varies sinusoidally with position on the surface. The elastic potential eliminates processes in which energy is exchanged between the particle and the surface, and the sinusoidal variation of speed makes the Liouvillian operator short ranged in reciprocal space. The calculation reduces to solving a kind of band-structure problem like that of the quantum motion of an independent electron in a crystal. In this Sec. the details of the model are presented, followed by the results of tridiagonalizing the Liouvillian, calculations of power spectra and diffusion constants by methods described in Sec. 2, and there is a discussion of the localization of the particle for slow speeds.

### 3.1 A Class of Corrugated Potentials

Because the potential is elastic, the sum of kinetic and potential energies of the particle remain constant, and for the purpose of solving Liouville's equation it is more convenient to specify the speed of the particle at each point on the surface rather than its kinetic energy or its total and potential energies. A sinusoidal variation of speed with position produces a Liouvillian of the shortest, non-zero range in function space, as can be seen below. Consider a square lattice of lattice constant $a$ and a family of potentials related by a parameter q for which the speed varies sinusoidally within the unit cell. For x and y coordinates in units of $a/(2\pi)$ let the speed s(x,y) be,

$$s(x, y) = \cos x + \cos y + q, \tag{15}$$

in units of $s_0$ which is half the difference between the speeds at the center of the cell and at the center of one side, for the cell $|x|<\pi$ and $|y|<\pi$.

The parameter q uniformly shifts the speed of the particle at various positions in the cell. Since the kinetic energy is proportional to the square of the speed, changing q changes the kinetic energy at each point quadratically, and hence also changes the potential quadratically to keep the total energy constant. As a result this class of systems does not correspond to a single potential with various total energies, but rather to a class of potentials, each with a particular total energy, for which size of the barrier between neighboring cells varies. For each value of q, the potential has a maximum along the contours of zero speed which the particle cannot cross. For q zero, the surface is divided into squares and the particle is trapped in which ever square it starts. For q between -2 and 2, but not zero, there is one region which spreads over all the cells and could support diffusion, and one finite region per cell in which the particle is trapped if it starts there. For |q| greater than 2, the particle can go anywhere on the surface, but is still scattered by variations in the potential.

This class of potentials allows the barrier between cells to be varied while preserving analyticity of the Liouvillian, at the cost of varying the potential along with the total energy. For a total energy less than the maximum potential, the kinetic enegy goes to zero linearly near a classical turning point, and so, the speed has a square-root singularitiy at the classical turning point. Since the Liouvillian depends linearly on the velocity, this produces a square-root singularity in the Liouvillian which complicates calculations. While singular Liouvillians can be treated by the methods described in this paper, it seems sensible to study non-singular cases first.

In systems of dimension greater than one, it is hard to see how the behavior of the speed near contours of zero speed can have any qualitative effect on the dynamics because only an infinitesimal proportion of trajectories sample this part of phase space. Only when the velocity is perpendicular to the contour of zero speed does the particle get close to the contour, and in that case only infinitely slowly. When the velocity is oblique to the contour, the particle never gets close and so is not affected by whether or not there is a singularity.

3.2 Liouville's Equation

The phase space for this model is three-dimensional and conveniently described by the coordinates x and y together with the angle $\theta$ between the direction of the velocity and the x-axis. For each choice of x, y, and $\theta$ the magnitude of the velocity is $\pm s(x,y)$, and so Liouville's equation relates the time-derivative of a function at each point in phase space to the sum of rates of change of each coordinate times the derivative of the function with respect to that coordinate. The Liouvillian operator for this system is,

$$L = -i\, s(x, y)\, [\cos\theta\, \partial/\partial x + \sin\theta\, \partial/\partial y] - i[\cos\theta\, \partial s(x, y)/\partial y - \sin\theta\, \partial s(x, y)/\partial x]\, \partial/\partial\theta, \quad (16)$$

where the coefficients of $\partial/\partial x$ and $\partial/\partial y$ are respectively the components of velocity in the x and y-directions, and the coefficient of $\partial/\partial\theta$ is the gradient of the speed in the direction of increasing $\theta$ which is perpendicular to the velocity.

The Liouvillian for this model is periodic in phase space because of the periodicity of the speed in the coordinates x and y and periodicity of the velocity in the angle $\theta$. Because of its periodicity, the Liouvillian is block diagonal in plane waves of x, y, and $\theta$, and only couples plane waves with wave-numbers differing by reciprocal lattice vectors whose wave-numbers are integers in each of the x, y, and $\theta$ directions. For a plane wave,

$$\phi(k_x, k_y, k_\theta) = \exp\{i\, (k_x\, x + k_y\, y + k_\theta\, \theta)\}, \quad (17)$$

the action of the Liouvillian operator is,

$$\begin{aligned}L\, \phi(k_x, k_y, k_\theta) = \{&(k_x - i\, k_y - k_\theta)\, \phi(k_x + 1, k_y, k_\theta + 1) + (k_x - i\, k_y + k_\theta)\, \phi(k_x - 1, k_y, k_\theta + 1) \\&+ (k_x + i\, k_y + k_\theta)\, \phi(k_x + 1, k_y, k_\theta - 1) + (k_x + i\, k_y - k_\theta)\, \phi(k_x - 1, k_y, k_\theta - 1) \\&+ (k_x - i\, k_y + i\, k_\theta)\, \phi(k_x, k_y + 1, k_\theta + 1) + (k_x - i\, k_y - i\, k_\theta)\, \phi(k_x, k_y - 1, k_\theta + 1) \\&+ (k_x + i\, k_y + i\, k_\theta)\, \phi(k_x, k_y + 1, k_\theta - 1) + (k_x + i\, k_y - i\, k_\theta)\, \phi(k_x, k_y - 1, k_\theta - 1) \\&+ 2\, q\, (k_x - i\, k_y)\, \phi(k_x, k_y, k_\theta + 1) + 2\, q\, (k_x + i\, k_y)\, \phi(k_x, k_y, k_\theta - 1)\}/4. \quad (18)\end{aligned}$$

Note that this representation of *L* as a matrix in terms of plane waves is Hermitian, so its spectrum is purely real. Although *L* has Bloch form, its diagonal matrix elements are zero while its off-diagonal matrix elements are proportional to the wave-number, in contrast to Hamiltonians for independent electrons in crystals which have diagonal matrix elements proportional to the square of the wave-number and off-diagonal matrix elements which depend on the differences of wave-numbers.

The advantage of this model in which speed varies sinusoidally across the unit cell is that *L* has non-zero matrix elements only between waves for which $k_x$, $k_y$, or $k_\theta$, changes by $\pm 1$. The off-diagonal matrix elements of the Liouvillian include the Fourier transform of the speed as a function of position, so if the speed were to have any singularities such as the square-root which arises from a kinetic energy going linearly to zero with the distance form the boundary of the classically allowed region, then the Fourier transform would be long-ranged in k-space and the off-diagonal matrix elements of the Liouvillian would decrease slowly with increasing differences in wave-numbers. But, because the speed in Eq. 15 goes linearly to zero with distance from a boundary, even trajectories headed directly toward the boundary only approach it exponentially slowly. Such trajectories which effectively stop are an infinitesimal fraction of all trajectories and therefor do not affect the results presented here.

3.3 Tridiagonalization of the Model

In order to calculate the diffusion constant for this model using Eq. 14, it is necessary to evaluate the projected resolvent for a plane wave in the limit as the wave-number k goes to zero. Taking $u_0$ to be $\phi(k_x, k_y, 0)$, and then using Eqs. 5, 6 and 18 to do the first step of the tridiagonalization gives,

$$b_0 = 2 \sqrt{[2 k^2 (1 + q^2)]}, \tag{19}$$

which goes to zero linearly with $k = \sqrt{(k_x^2 + k_y^2)}$. In the limit where k goes to zero along the $k_x$ axis,

$$u_1 = \{ \phi(1, 0, 1) + \phi(-1, 0, 1) + \phi(1, 0, -1) + \phi(-1, 0, -1) + \phi(0, 1, 1) + \phi(0, -1, 1)$$

$$+ \phi(0, 1, -1) + \phi(0, -1, -1) + 2 q \phi(0, 0, 1) + 2 q \phi(0, 0, -1) \} / [2 \sqrt{(2 + 2 q^2)}], \tag{20}$$

and this is the starting point of the tridiagonalization using Eqs. 5 and 6 in which the limit of k going to zero is implicit. Note that as a result of the gauge symmetry introduced in this formulation of classical mechanics, the relative phases of different components of $u_1$ depend on the direction from which k approaches zero, but the resulting physics does not.

Before presenting the results of tridiagonalization, there are several features of the Liouvillian to be noted. The first is that in the plane wave basis, the diagonal elements of the Liouvillian are all zero. Then, because the reciprocal lattice is cubic, $u_n$ for odd and even n lie of distinct sublattices, consistent with Sec. 2.2 the $\{a_n\}$ are all zero. The next point is that the sinusoidal part of s(x,y) couples each wave to eight nearby waves which belong to the same Body Centered Cubic sublattice. There are two disjoint Body Centered Cubic sublattices of the cubic lattice. It is the constant term in s(x,y) which couples waves in one sublattice to waves in the other, and these interactions between sublattices produces oscillations in the tridiagonalization when the coupling is sufficiently strong. Taking advantage of this simple structure of the Liouvillian, it is relatively easy to calculate the first 256 of the continued fraction coefficients.

Because the matrix elements of *L* are proportional to wave-numbers, the tridiagonal matrix elements $\{b_n\}$ approach a linear dependence on n for n large, as was seen for simpler models in previous work [1]. The recursion index n can be thought of as a wave-number associated with orthogonality of the $\{u_n\}$ and so the approach of the $\{b_n\}$ to a linear dependence on n is consistent with the linear dependence of matrix elements of Liouvillian on wave-number. Figure 2, shows how the $\{b_n\}$ deviate from linearity for several values of the parameter q in the model, in units of frequency f which is the unit of speed divided by the unit of distance in which x and y are measured. For all values of q, the deviations of the $b_n$ from linearity alternate in sign, but for large |q| there are oscillations superimposed on the alternation which are absent at low |q|. This qualitative change in the asymptotic behavior of the $\{b_n\}$ suggests a phase transition for some small value of |q|.

The asymptotic behavior of the $\{b_n\}$ can be analyzed by fitting their deviation from linearity to a combination of constant and oscillating functions of n. First the linear behavior is subtracted as in Fig. 2, then the deviations are multiplied by alternating signs to remove the odd-even alternation. A sinusoidal oscillation is then fit to the remainder using its first and second differences to remove the constant term which is then fit to what is left after the oscillation has been removed. The wave-number of the oscillation of $b_n$ in n is plotted in Fig. 3 for different values of the parameter q in the model. It is clear that there is a singularity in the dependence of this wave-number on q near 0.3, and the implications of this are discussed below. There also appear to be weaker singularities in the wave-number for values of q near 0.55 and 0.5; these may be due to the small number of fitting parameters, or some phase transitions not yet identified.

3.4 Power Spectra, Diffusion, and Classical Localization

As discussed in Sec. 2, the resolvent connects the model described by some Liouvillian with observations of the physical system being modeled. Using Eq. 11, the tridiagonalizations presented in the previous Subsec. allow calculations of resolvents projected on the function $u_1$ in Eq. 20, for the model Liouvillian with various values of q. The phase-space function $u_1$ is periodic but peaked toward the center of each unit cell. Its resolvent describes the evolution of that function, and the imaginary part of the resolvent is proportional to the power spectrum of fluctuations in the autocorrelation or overlap of $u_1$ at time zero with its evolution at time t. Figure 4 shows parts of the power spectra for values of q used in Fig. 2, averaged by replacing the infinite tail of the continued fraction with $i/b_{256}$. Note that as |q| increases, the projected power density near zero frequency increases, consistent with increasing diffusion.

However, the averaged power spectra in Fig. 4 do not reflect the long-time properties of the system, which are contained in the singularities of the spectra and are determined by the asymptotic behavior of the tridiagonalization. While it is not possible to calculate all the $\{b_n\}$, those calculated seem to settle into stable patterns by about $b_{100}$, and in what follows, the asymptotic behavior of the $\{b_n\}$ is extrapolated from the $\{b_n\}$, for n between 200 and 256, by fitting them to a form which is the sum of terms which are linear and exponential in n. The exponential terms alternate in sign for n odd and even, and consist of either a decreasing real exponential or a linear combination of sine and cosine. These fits are good enough that no join is visible when plots such as those in Fig. 2 are extended beyond $b_{256}$ using the fits.

The diffusion constant depends on $R_1(0)$, so the fitted asymptotics are applied to this first. Since $\omega$ is zero, the continued fraction expansion for $R_1(0)$ can be converted into an infinite product,

$$-(b_2 b_4 \ldots b_{2n})^2 / [(b_1 b_3 \ldots b_{2n-1})^2 b_{2n+1}] < R_1(0) < (b_2 b_4 \ldots b_{2n})^2 b_{2n+2} / (b_1 b_3 \ldots b_{2n+1})^2 \quad (21).$$

If these bounds converge, then $R_1(0)$ is zero, but if they diverge, $R_1(0)$ is singular, and if they neither diverge nor converge, the eigenfunctions of the Liouvillian at $\omega=0$ form a time-reversal doublet so that $R_1(0)$ is determined by boundary condition at $n=\infty$. For each extension of the continued fraction by a pair of levels, the upper bound is multiplied by the factor of $b_{2n}b_{2n+2}/(b_{2n+1})^2$, and the lower bound is multiplied by the factor $(b_{2n})^2/[b_{2n-1}b_{2n+1}]$, so convergence of the bounds depends on how these factors compare with unity.

For |q| less than about 0.3, where the kink occurs in Fig. 3, the odd numbered $b_n$ lie above the average linear growth, and the even numbered $b_n$ lie below the average linear growth, by an amount which seems to become constant for large n, as for example in Fig. 2(a). For such $\{b_n\}$ when deviation from linear growth is small,

$$(b_{2n})^2 / [b_{2n-1} b_{2n+1}] \approx b_{2n} b_{2n+2} / (b_{2n+1})^2 \approx 1 - (b_{2n+1} + b_{2n-1} - 2b_{2n}) / b_{2n} \qquad (22)$$

to order $1/n$. So as long as $b_{2n+1}+b_{2n-1}-2b_{2n}$ exceeds some positive number for large n, the two bounds on $R_1(0)$ converge to zero and there is no diffusion. If as happens for $|q|$ greater than 0.3, the odd and even $\{b_n\}$ oscillate from one side to the other of the average linear growth, as for example in Fig. 2(c), then the bounds on $R_1(0)$ also oscillate and there is diffusion.

    The tridiagonalization and asymptotic analysis of the model can now be used to characterize the transition between non-diffusive and diffusive regimes as q varies in the model. From Eq. 14 the diffusion constant depends on two quantities, $\beta^2$ and $R_1(0)$. For this model, $\beta^2$ is $8(1+q^2)$ according to Eq. 19, so the crucial dependence of the diffusion constant on q comes from $R_1(0)$. For $|q|$ less than about 0.3, $R_1(0)$ converges to zero, so the diffusion constant is zero. For $|q|$ greater than about 0.3, the wave-number in n of the oscillations of the $\{b_n\}$ appears to rise linearly with q. When the $\{b_n\}$ oscillate about linearity, the products in Eq. 21 are truncated to the period of the oscillation which, according to Fig. 3 is approximately proportional to $1/(q-q_c)$ where $q_c$ is the critical value of q. This gives $R_1(0)$ a power law dependence on $q-q_c$, where the power is proportional to $(b_{2n+1}+b_{2n-1}-2b_{2n})/b_{2n}$, which depends on the details of the potential in the model.

    The key to interpreting this transition in the diffusion constant is that for q=0, the particle is trapped in a single cell. As $|q|$ varies away from zero, the tridiagonalization does not change qualitatively until about 0.3, suggesting that the states between zero and 0.3 are also localized; probably not in a single cell, but in some finite region of the surface. Since the particle is free to move from one cell to the next for these values of q, this localization appears to violate any assumption that the particle's motion in different cells is unrelated, and more specifically seems to violate any relation between the diffusion constant and the escape time from a single cell [15].

4. Weak Inelastic Scattering and Surface Diffusion

The model analyzed in the previous Sec. is simple because the interaction between the particle and the surface is purely elastic. The problem with the elastic model is that an adatom can be trapped forever if it does not have enough kinetic energy to get over the barriers around it, or if its trajectory never happens to go through a gap in some barrier. However, a real adatom cannot be trapped forever because eventually there is a thermal fluctuation large enough to get it over any barrier, or change its velocity enough to get it through a gap. The thermal fluctuations of the adatom are due to inelastic processes by which it transfers energy to or from the surface and makes transitions from one state of the elastic potential to another. If the elastic motion of the adatom is already diffusive, then inelastic processes can increase a small diffusion constant or reduce a large one. However, if the adatom is in one localized state of the elastic potential, then an inelastic process can transfer the adatom to a different localized elastic state. Since the entire surface is covered by elastic states, even if they are all localized, inelastic processes restore diffusion by sending the adatom on a random walk from one localized elastic state to another.

One approximation for inelastic processes is to replace them with a random force so that the equation of motion for the adsorbed atom becomes a stochastic differential equation. Risken and others, see Ref. [8], have developed analytic and numerical methods for solving such equations. Using these methods, Costantini and Marchesoni [16] have investigated a transition at zero temperature between locked and running states as the tilt is varied for damped motion of particles in a tilted sinusoidal potential. This model of a tilted potential is relevant because, like the model in Sec. 3, the transition takes place at a value of the tilt significantly larger than the minimum necessary to overcome the barrier between neighboring cells. What remains mysterious is whether there is more than a superficial connection between these two transitions.

A precise calculation of the effects of inelastic processes would require the addition of all the surface degrees of freedom to the model used in the previous Sec. While this is not out of the question, weak inelastic effects can be included much more easily as a perturbation of the purely elastic model. This approach to weak inelastic effects for classical systems is similar in spirit to the way inelastic scattering is included in quantum models to produce what is called, 'variable range hopping,' see Mott and Davies [17]. The idea is that the adatom propagates elastically between inelastic events which are approximated by instantaneous changes in the velocity of the particle. The effect of these inelastic events is to transfer the particle from one elastic state to another which, of course, includes transfer between diffusing states, between diffusing and localized states, as well as between localized states.

The purpose of this Sec. is to add weak inelastic effects to the elastic model discussed above, and to estimate the temperature dependence of surface diffusion constants. The first problem is that while the parameter q in the above model changes the contours of zero speed, it also changes the elastic potential because the contours of zero speed are maxima of the potential. In order to include inelastic effects as a perturbation of the elastic motion, the model must have a single potential in which the particle moves whatever its energy. The contours of zero speed for different q can be converted to contours of zero speed for particles with different total energies in a single potential by giving up the linear relation between speed and distance from a contour of zero speed. The potential with the same contours of zero speed as the model in Eq. 15 is,

$$V(x, y) = 2 - \cos x - \cos y \tag{23}$$

where V is the potential energy in units of the mass of the particle times $s_0^2$. The contours of zero speed are where V(x,y) is equal to the total energy of the particle, and these contours coincide with those of Eq. 15. For each value of the particle's total energy, the only difference between the new and old models is that the speed goes to zero as the root of the distance from a contour of zero speed in the new model, and it goes linearly with distance in the old model. The parts of the surface accessible to the particle are the same in both cases, only the probability of finding the particle close to a contour of zero speed is different. The assumption required to add inelastic effects to the results of Sec. 3 in this simple way, is that only the shapes of the contours of zero speed matter, specifically that the way the particle approaches a contour of zero speed does not matter despite its effect on the range of the Liouvillian in reciprocal space - see Sec. 3.2.

The simplest effect of classical localization is to increase the effective barrier for diffusion. If the particle only had to get over the potential barrier between cells in order to diffuse, its diffusion constant would be proportional to exp{-2/T} where 2 is the height of this barrier for the potential in Eq. 23 and the temperature T is measured in the same units of energy. Within the assumption that it is only the shape of the contours of zero speed which matter, then the effect of localization on this model is to make the diffusion constant proportional to exp{-2.3/T} where the increase on 0.3 in energy is necessary to make the contours of zero speed in Eq. 23 coincide with those at the critical velocity for the model in Sec. 3. This shift in the barrier height is only a practical test of the theory presented here if the potential barriers are known for a specific system. These barriers are difficult to measure independently from the diffusion constant.

A subtler effect of the localization is a phenomena known in electronic transport as variable-range hopping, and it affects the electronic conductivity at temperatures below those for which the conduction is activated [17]. For an atom diffusing on a surface, the idea is that inelastic processes need not raise the energy of the atom to that of diffusing states, but can produce diffusion by simply transferring the atom from one localized state to another. If the atom is weakly coupled to the surface at temperature T, then the atom will have a Boltzmann distribution of total energies $\Delta$, which is $\exp\{-\Delta/T\}$. For T small compared to 0.3 which is the difference between the critical total energy for diffusion 2.3 and the height of the potential barrier 2, diffusion is dominated by transitions between the localized elastic states. The activated diffusion of the previous paragraph occurs at temperatures comparable with the activation energy 2.3.

The Boltzmann distribution of total energies is not sufficient to determine the diffusion constant because a distance is needed. For the localized state with total energy $\Delta$, this is the localization length $d(\Delta)$ which has not yet been determined for the model discussed above. The contribution to diffusion from the elastic state at energy $\Delta$ is $d(\Delta)\exp\{-\Delta/T\}$ which for a given temperature T has a maximum at some energy E(T) for which the elastic state has a localization length $L(T)=d(E(T))$. Approximating diffusion of the adatom as hops of this most probably length gives a diffusion constant which is proportional to $L(T)^2\exp\{-E(T)/T\}$ which has a temperature-dependence characteristic of the energy-dependence of the localization length.

Acknowledgments

The Author thanks Hannes Jonsson for encouraging this work surface diffusion, John Hannay for illuminating discussions, and Peter Feibelman for the opportunity to present preliminary results. Some of the numerical calculations were performed on computers provided through the Computational Science Institute of the University of Oregon by the National Science Foundation's Office of Science and Technology Infrastructure under grant no. STI-9413532, and other numerical calculations were performed on the Hitachi S3600 located at the University of Cambridge High Performance Computing Facility.

**Figure Captions**

1. Contours of zero speed for a particle having speeds 0 (-), 0.5 (o), and 1 (+) at the saddle points (±0.5, 0) and (0, ±0.5). (Note that the X and Y-coordinates are plotted in units of the lattice constant $a$.)

2. Differences between $b_n$ and a linear fit for speeds 0 (a), 0.5 (b), 1.0 (c), 2.0 (d), and 3.0 (e).

3. Wave-number of the oscillatory component of $b_n$ for different speeds.

4. Averaged projected power spectra, $|\text{Im}\{R_1(\omega)\}|$ averaged of an interval of $\omega$, for initial speeds 0 (a), 0.5 (b), 1.0 (c), 2.0 (d), and 3.0 (e)

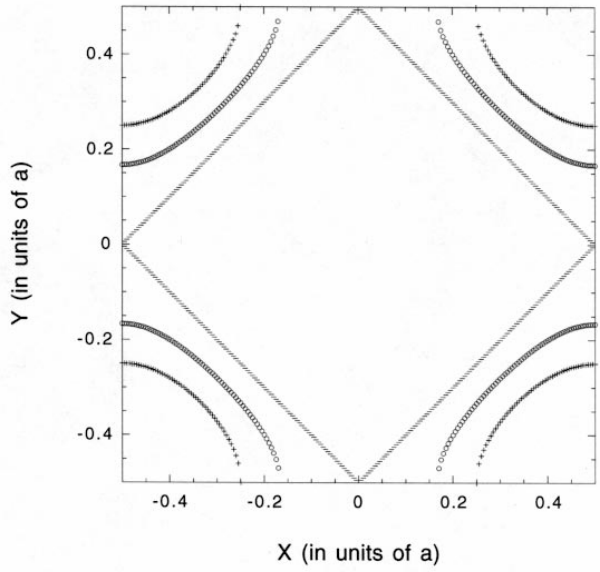

Figure 1.

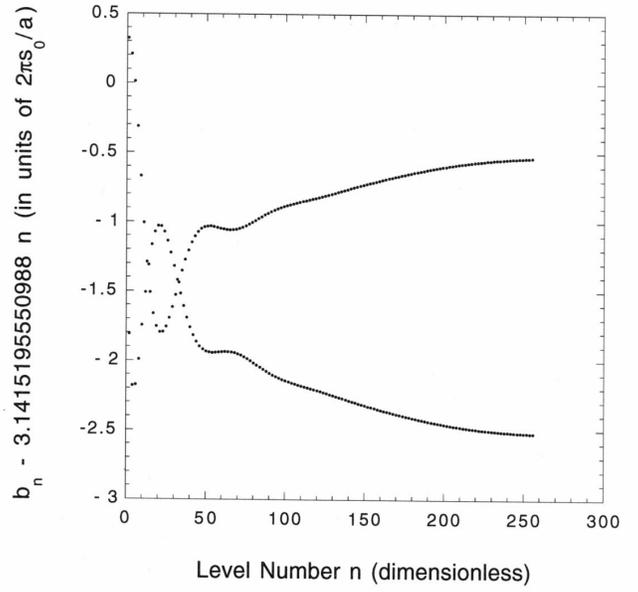

Figure 2(a).

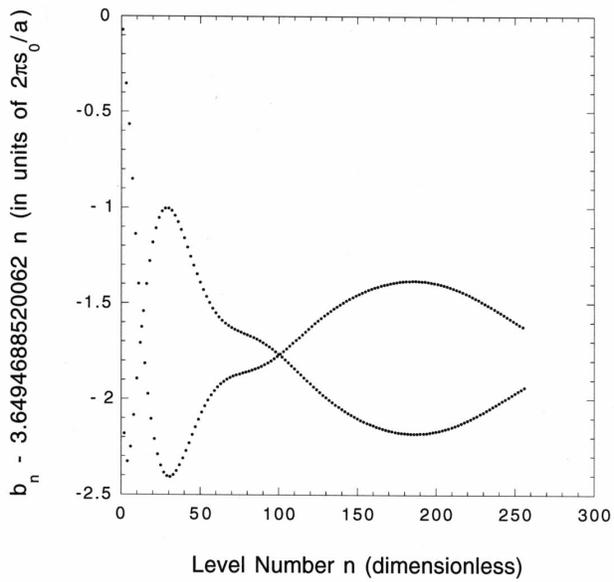

Figure 2(b).

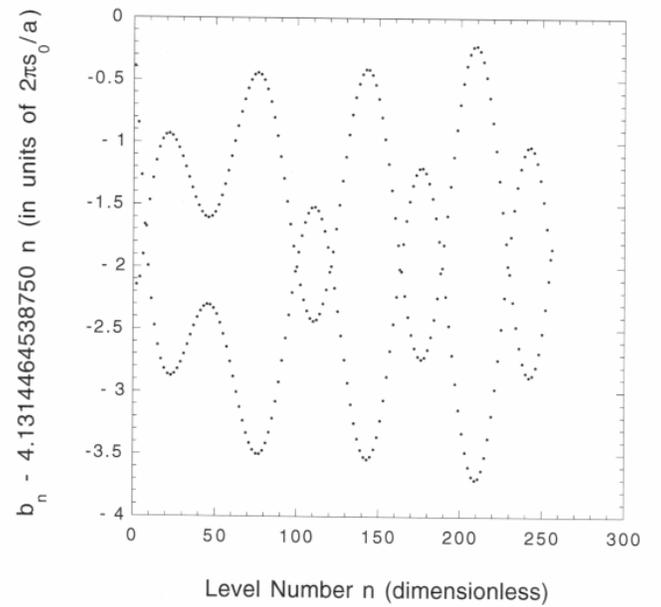

Figure 2(c).

Figure 2(d).

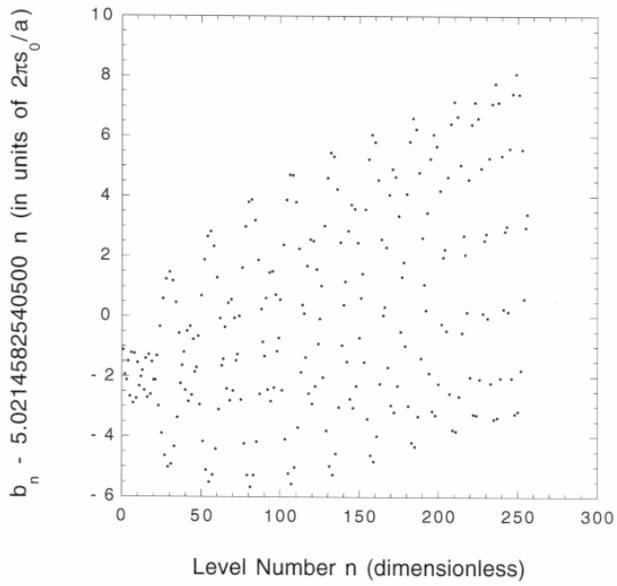

Figure 2(e).

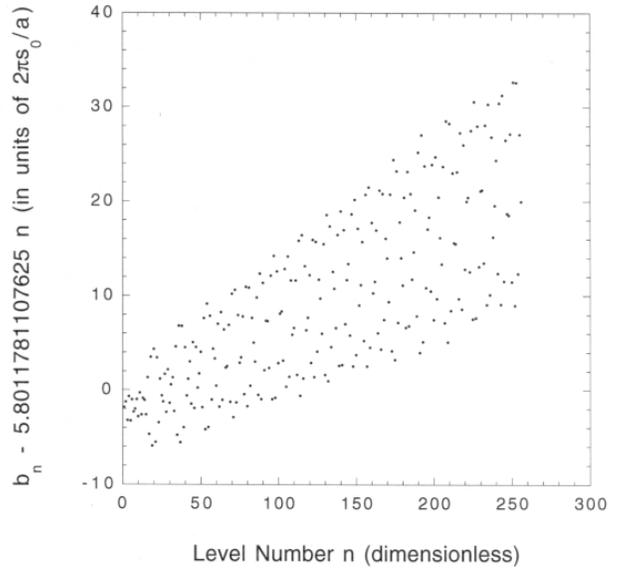

Figure 3.

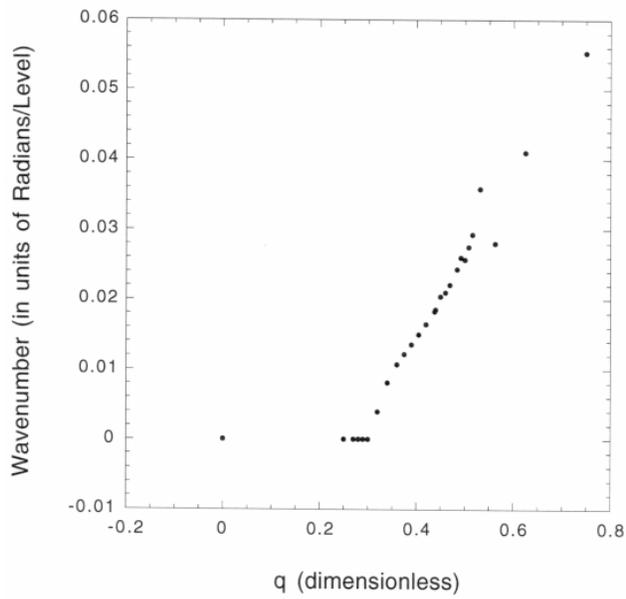

Figure 4(a).

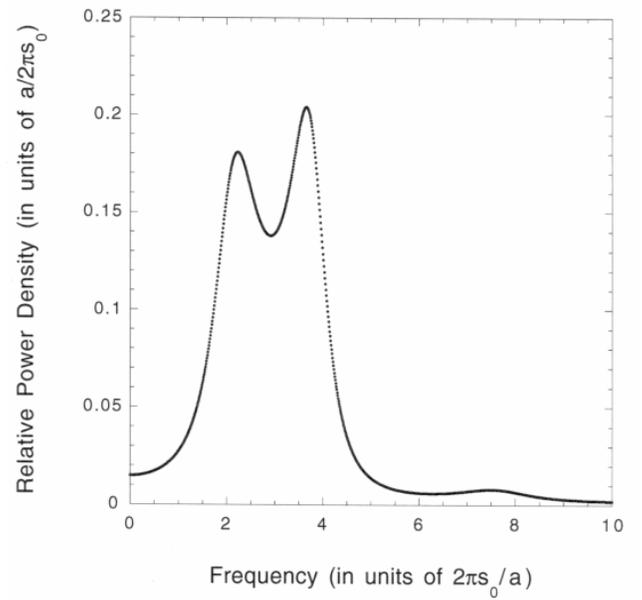

Figure 4(b).

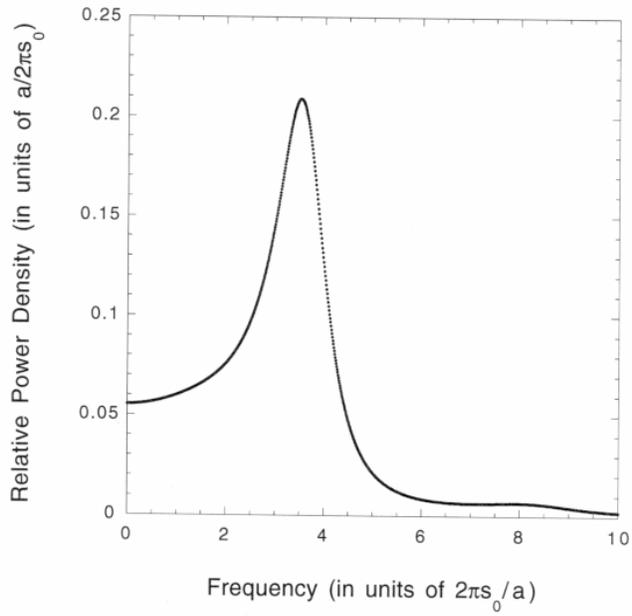

Figure 4(c).

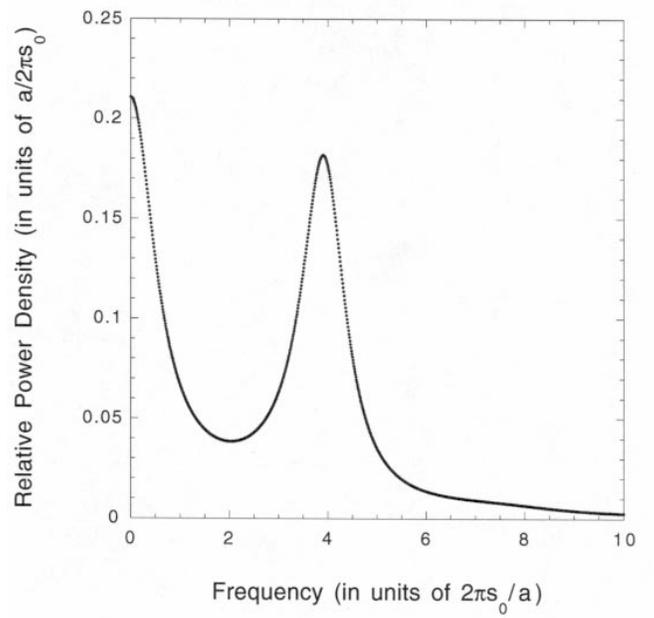

Figure 4(d).

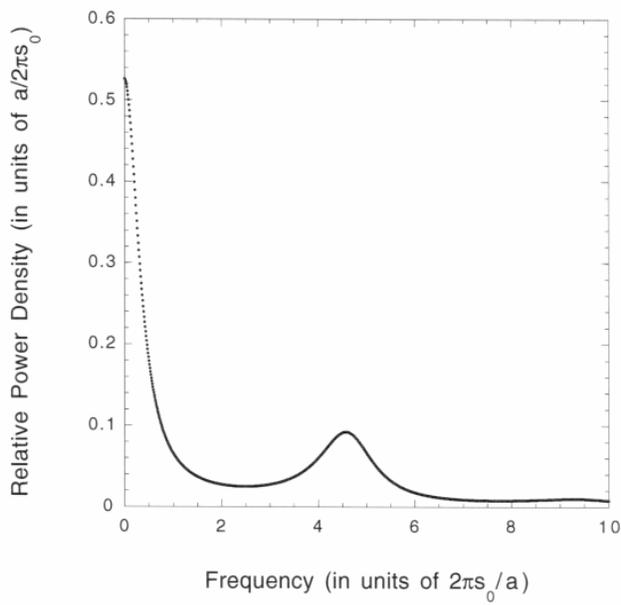

Figure 4(e).

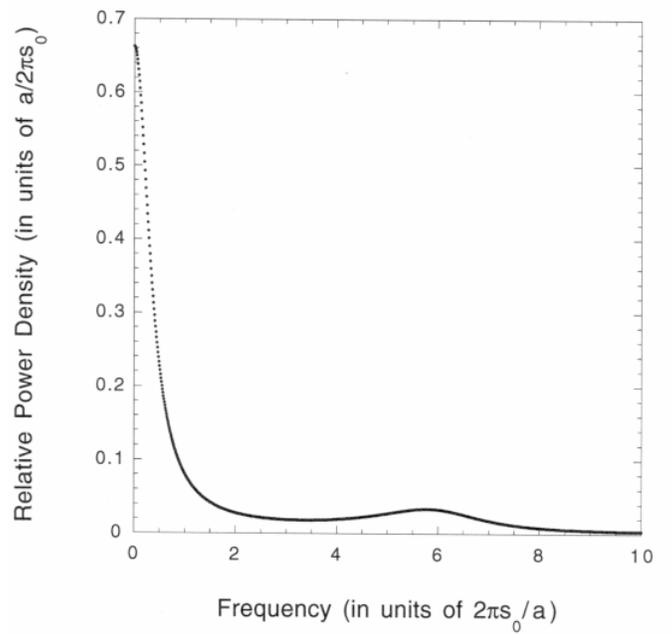